\documentclass[11pt, oneside]{article}   	
\usepackage[margin=1in]{geometry}               
\usepackage{graphicx}				
										
\usepackage{amssymb}
\usepackage{physics}
\usepackage{amsmath}
\usepackage[dvips]{epsfig}

\usepackage{amsfonts}
\usepackage{subfig}

\usepackage{mathrsfs}

\usepackage{xcolor}

\usepackage{bm}

\usepackage{blindtext}

\usepackage[title]{appendix}

\usepackage{cite}

\usepackage{authblk}

\def\cchi{\raise2pt\hbox{$\chi$}} 

\title{\bf{Qubit Lattice Algorithm Simulations of the Scattering of a Bounded Two Dimensional 
Electromagnetic Pulse from an Infinite Planar Dielectric Interface}}

\author {Min Soe ${}^1$, George Vahala ${ }^{2}$, Linda Vahala ${ }^{3}$, Efstratios Koukoutsis ${ }^{4}$, Abhay K. Ram ${ }^{5}$, Kyriakos Hizanidis ${ }^{4}$\\
${ }^{1}$ Department of Mathematics and Physical Sciences, Rogers State University, Claremore,OK 74017\\
${ }^{2}$ Department of Physics, William \& Mary, Williamsburg, VA23185 (retired) \\
${ }^{3}$ Department of Electrical \& Computer Engineering, Old Dominion University, Norfolk, VA 23529\\
${ }^{4}$ School of Electrical and Computer Engineering, National Technical University of Athens,Zographou 15780, Greece \\
${ }^{5}$ Plasma Science and Fusion Center, MIT, Cambridge, MA 02139\\}
\date{}

\begin{document}
\maketitle
\begin{abstract}
Qubit lattice algorithm (QLA) simulations are performed for a two-dimensional (2D) spatially bounded pulse propagating onto a plane
interface between two dielectric slabs.
QLA is an initial value scheme that
consists of a sequence of unitary collision and streaming operators, with appropriate potential operators, that recover
Maxwell equations in inhomogeneous dielectric media to second order in the lattice discreteness.
For the case of total internal reflection, there is transient energy transfer into the second medium 
due to the evanescent fields as the Poynting unit vector
of the pulse is rotated from its incident to reflected direction.  Because of the finite spatial extent of the pulse, a self-
consistent Goos-Hanchen-type displacement along the interface is found without imposing any explicit interface boundary
conditions on the fields.
For normal incidence.  the standard  Fresnel coefficients are recovered for appropriately averaged QLA fields.
Energy is conserved at all times to seven significant figures.

\end{abstract}

\section{Introduction}

\qquad Recently, we have been developing a qubit lattice algorithm (QLA) for the solution of Maxwell equations in dielectric media [1-6].
In principle, QLA is a quantum-inspired algorithm built from an interleaved sequence of unitary
collision and streaming operators that act on a lattice representation of Maxwell equations, recovering
the continuum dynamics to second order in the lattice spacing.  QLA is not a direct finite difference representation of
the field equations themselves.
As a quantum algorithm, the
QLA representation is ideal for direct encoding on a quantum computer [7-13].  Moreover, it 
exhibits ideal parallelization to all available cores on classical supercomputers. 
This efficiency arises because the collision operator acts purely
locally at each lattice site, while the streaming operator simply shifts data to neighboring sites.
When, spatial gradients in the dielectric medium are present such as interfaces or inhomogeneous
regions, additional potential operators have to be introduced into the collision-streaming sequence.
These operators are typically non-unitary but sparse.

Previous QLA simulations [1-4] successfully considered the scattering of a 1D Gaussian pulse from localized dielectric structures. However, for non-local dielectrics the QLA studies were restricted to normal incidence because periodic boundary conditions are applied at the lattice boundaries.   It is not straightforward to investigate oblique incidence using this 1D pulse scattering from infinite plasma slab interfaces.
 In addition, it was determined that for 1D QLA runs to rigorously conserve energy, an asymptotic scaling of the transition region between the two dielectrics is required. Thus, QLA simulations for initial 1D pulses did not permit us to study phenomena like total internal reflection and the Fresnel relations for reflected and transmitted fields. 
 Nevertheless,  the study of these incident 1D pulses interacting with small localized 2D and 3D dielectric objects yielded very interesting results [4] on the transient effects of reflections and transmissions from within the dielectric object itself while 
rigorously conserving the energy.

In the present paper, we consider a QLA for a 2D bounded pulse, where the non-zero domain
of the pulse has finite measure in both spatial coordinates, Fig. 1. Specifically, we consider an incident bounded 2D electromagnetic TEM pulse onto a plane interface between two scalar dielectrics with refractive indexes $n_1$
and $n_2$.  
The plane of incidence is defined by the standard Cartesian $(x,y)$ coordinates with the two dielectric regions ${0 \le x < L, 0 \le y < L/2}$
and ${0 \le x < L, L/2 \le y < L}$.  For $p$-polarization [8], the incident electric field, $\mathbf{E}$, is in the $x-y$ plane and the corresponding
magnetic field, $\mathbf{H}$, is in the $-z$-direction.
For the initial 2D pulse we use its natural coordinate system - a rotated Cartesian system $(\zeta, \chi)$ where the direction of incident
propagation is $\mathbf{\hat{\zeta}}$, the magnetic field is in the $-\hat{z}$ direction, and the  electric field in the $\hat{\chi}$
direction.  Thus the incident magnetic field amplitude is (Fig. 1)
\begin{equation} \label{eq1}
H_z(\zeta,\chi) = - Exp \left[- \left( \frac{\zeta - \zeta_0}{\zeta_w} \right)^2  -  \left( \frac{\chi - \chi_0}{\chi_w} \right)^2  \right]  cos(k \zeta) .
\end{equation}
Here $\zeta_w$ defines the extent of the packet in the $\zeta$-direction, and similarly $\chi_w$ in the $\chi$-direction. 
The two coordinate systems are related by the rotation matrix
 \begin{equation}\label{coordss}
\begin{bmatrix}
x \\
y
\end{bmatrix}=\begin{bmatrix}
cos \,\theta & sin \, \theta \\
-sin \, \theta & cos \,\theta  
\end{bmatrix}\begin{bmatrix}
\zeta \\
\chi
\end{bmatrix}
\end{equation}
where $\theta$ is the angle of incidence in the $(x,y)$-plane.

 The electric field amplitude $E_{\chi}(\zeta,\chi)$ is related to the magnetic field amplitude $H_z$
by the impedance of the medium, $Z=E_{\chi}/H_z$.

 \begin{figure}[!h!p!b!t] \ 
\begin{center}
\includegraphics[width=4.1in]{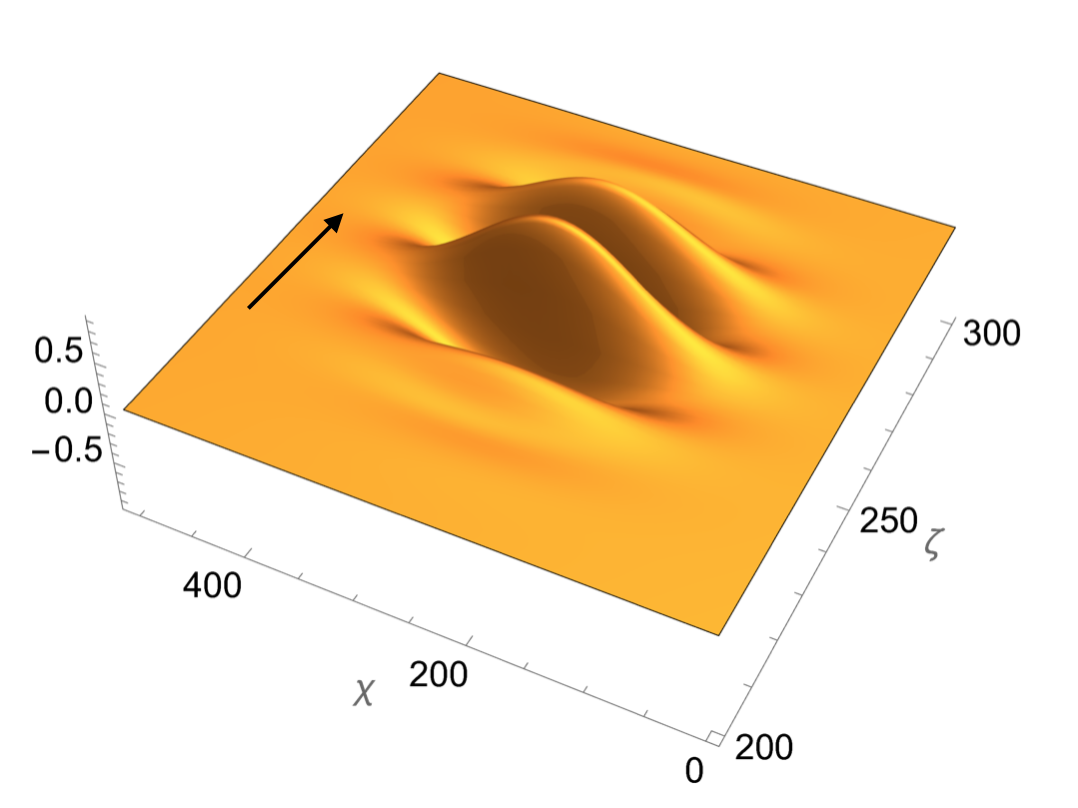}
\caption{ \it{Initial 2D magnetic field amplitude $H_z$ that is bounded in both $\zeta$ and $\chi$.  The initial pulse propagates in the $\zeta$-direction in
dielectric $n_1$, with dispersion relation $\omega / k = c/n_1$.
The initial electric field amplitude $E_{\chi} = Z_1 H_z$.  The square lattice has a spatial grid of side $L = 1024$.}
}
\end{center}
\end{figure}

In Section 2, QLA simulations are presented which yield transient effects in total internal reflection of the pulse
as well as the recovery of the spatial Goos-Hanchen [14] shift of the pulse along the interface.
Details of the transient fields generated around 
the interface region are presented as the incident pulse is steered into the reflected pulse. 
The Fresnel conditions for normal incidence are considered in Section 3.   In all these QLA simulations, total
energy is conserved to at least 7 significant figures.
 For completeness we briefly summarize
the QLA operators required to recover the Maxwell equations in the Appendix as well as some quantum information
science ideas on how to deal with non-unitary operators for quantum computing.

\section{Total Internal Reflection}
Consider the electromagnetic propagation of this bounded pulse, Fig. 1, onto a slab dielectric interface at $y=L/2$.  
For $y < L/2$, the dielectric slab has refractive
index $n_1=2$ while the dielectric slab $y \ge L/2$ has refractive index $n_2=1$.  The critical angle for total internal reflection is $\theta_c = 30^o$ [8].  For
p-polarization, we first plot the time evolution of the
 transient energy in each dielectric slab when the angle of incidence 
$\theta  = 25^o < \theta_c$ and when $\theta  = 35^o > \theta_c$.
The time-dependent electromagnetic energy in regions $y<L/2$ and $y \ge L/2$ is 
\begin{equation} \label{eq2}
\begin{aligned}
\mathcal{E}_1(t) =  \int_0^{L/2-1} dy \int_0^L dx  \left[\epsilon_0 n_1^2 \mathbf{E}_1^2 + \mu_0  \mathbf{H}_1^2 \right]  \, , \quad
\mathcal{E}_2(t) =  \int_{L/2}^L dy \int_0^L dx \left[\epsilon_0 n_2^2 \mathbf{E}_2^2 + \mu_0\mathbf{H}_2^2 \right]  .
\end{aligned}
\end{equation}  

 \begin{figure}[!h!p!b!t] \ 
 \begin{center}
\includegraphics[width=4.5in]{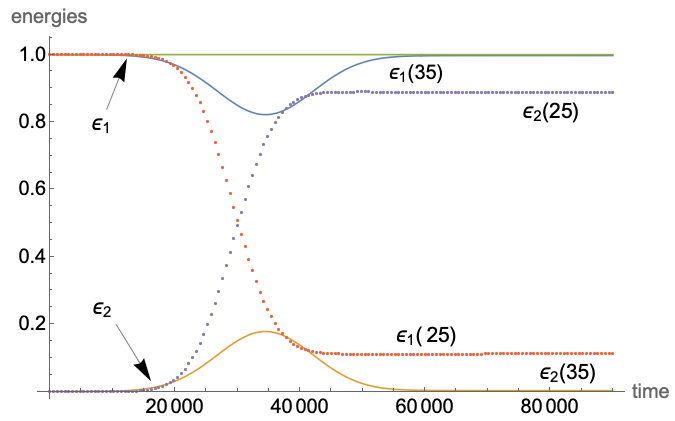}
\caption{ \it{The time evolution of the normalized energy in dielectric $n_1$, 
$\epsilon_1(t) = \mathcal{E}_1(t) / [\mathcal{E}_1(0) + \mathcal{E}_2(0) ] $, 
and in dielectric $n_2$, $\epsilon_2(t) = \mathcal{E}_2(t) / [\mathcal{E}_1(0) + \mathcal{E}_2(0) ] $
for $\theta = 25^o < \theta_c$ - - - dashed curves, and for $\theta=35^o>\theta_c$ - solid curves.  At every time output, the total energy,
$\mathcal{E}_1(t) + \mathcal{E}_2(t) = const. $ to the 7th significant figure.  In these simulations $L=1024$.
Thus the $\epsilon_1(35)$-plot is the time evolution of the  normalized energy in the refractive index medium $n_1=2$ 
for angle of incidence $\theta = 35^o$.
}
}
\end{center}
\end{figure}
\noindent In $QLA$-units, the simulation time $t$ is normalized by a factor $1/c$, where $c$ is the speed of light in vacuum.
From Fig. 2, the pulse for time $t< 18 k$ remains within the dielectric $n_1$, so that $\mathcal{E}_2(t) = 0$. 
It is important to note that since QLA is an initial value algorithm, no boundary or jump conditions are applied at $y=L/2$.  

\subsection*{Case 1: $\theta = 25^o < \theta_c$ }
Because the pulse has a finite width, there is a finite interaction time with the dielectric
interface during which there is a smooth temporal transition as it 
splits into transmitted and reflected components, as shown in the energy profiles, $\epsilon_2(25)$ and
$\epsilon_1(25)$, of Fig. 2.
For $t > 42K$ these pulses then propagate undistorted in their respective dielectric regions.

\subsection*{Case 2: $\theta = 35^o > \theta_c$ }
In our $QLA$-simulations for  $\theta = 35^o > \theta_c$, we observe 
total internal reflection for $t > 60k$, with normalized energy $\epsilon_2(35) = 0$ and $\epsilon_1(35) = 1$  , Fig. 2.  
During the intermediate interaction time-window $18k < t < 60K$, the finite
pulse width leads to transient evanescent fields in the
second dielectric $n_2$ associated with non-propagating energy transfer which nevertheless contributes to
the reflected pulse in the $n_1$ medium, along with a Goos-Hanchen [14]-like spatial shift along the
interface. The time evolution of the $H_z > 0$ part of the profile is presented in Figs 3 and 4.

For $t \approx 30k$, Fig 3(e), there is a symmetry about the interface normal in the structure of the fields together with
a peak 
in the transient total energy in the dielectric medium $n_2$, all localized near the interface $y = L/2$, Fig. 2

 \begin{figure}[!h!p!b!t] \ 
\begin{center}
\includegraphics[width=3.1in]{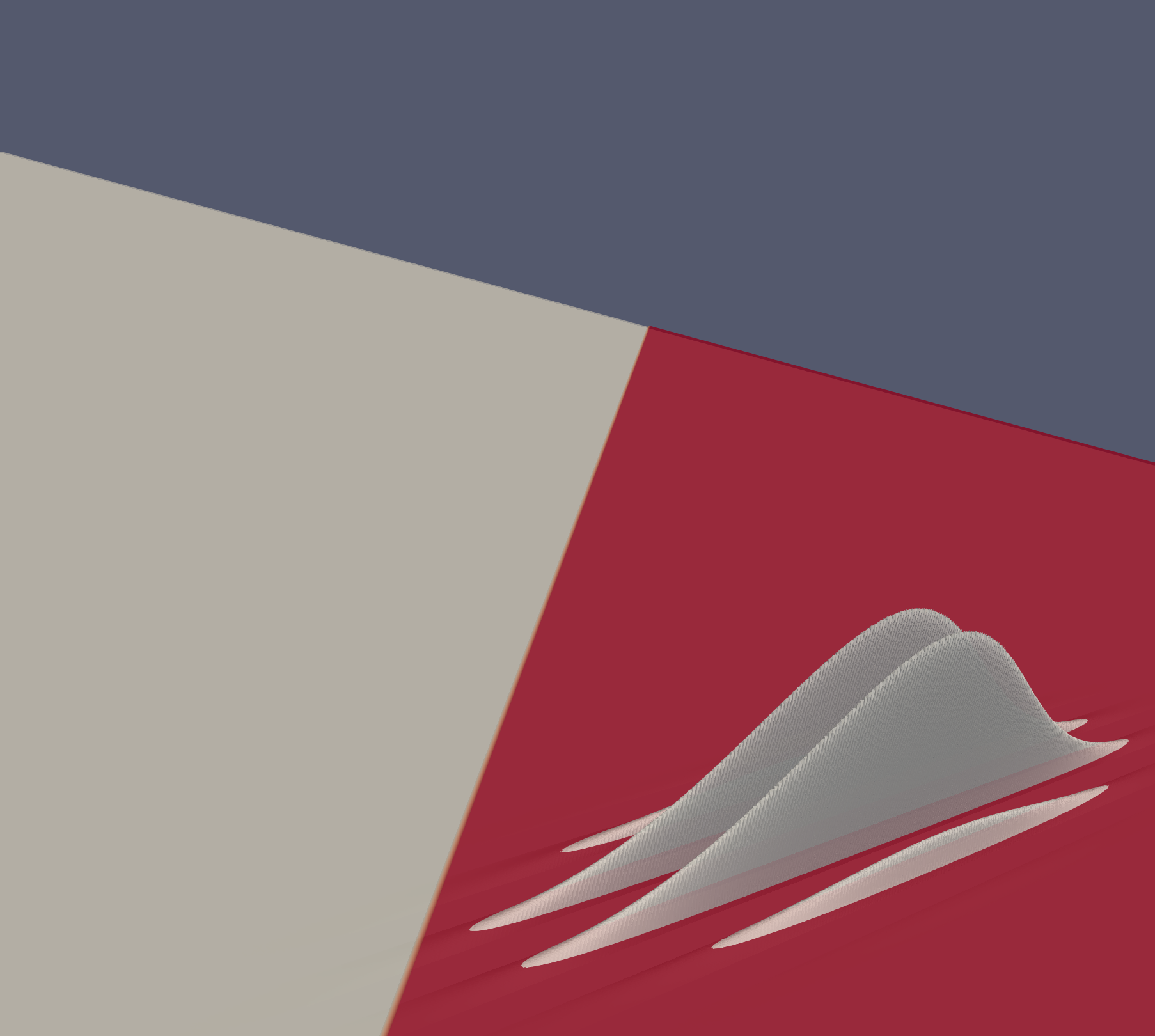}
\includegraphics[width=3.1in]{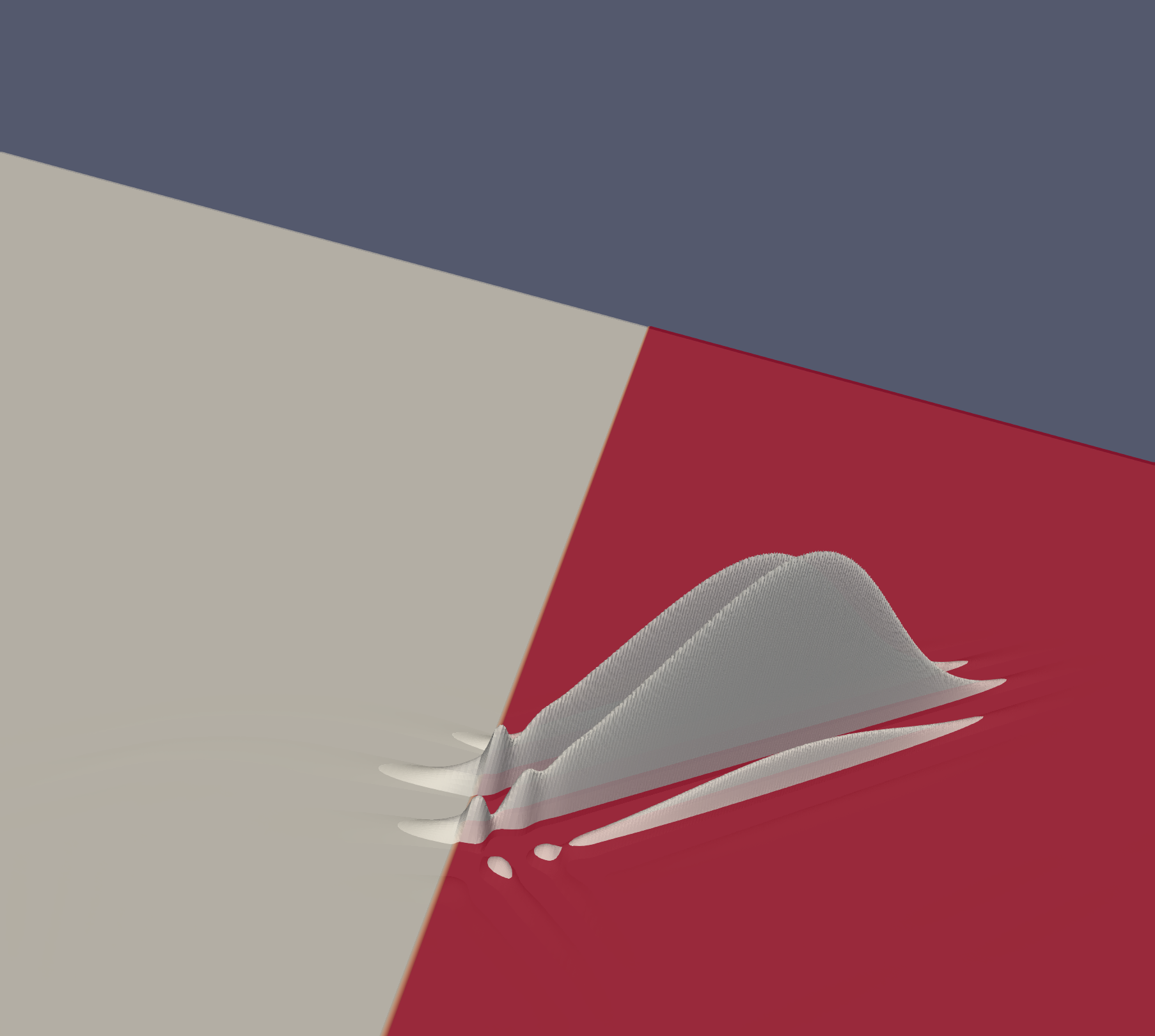}
.   (a) $H_z > 0 $ at $t= 0k$    \qquad   \qquad  \qquad  \qquad (b)  $H_z > 0$ at $t = 12k$
\includegraphics[width=3.1in]{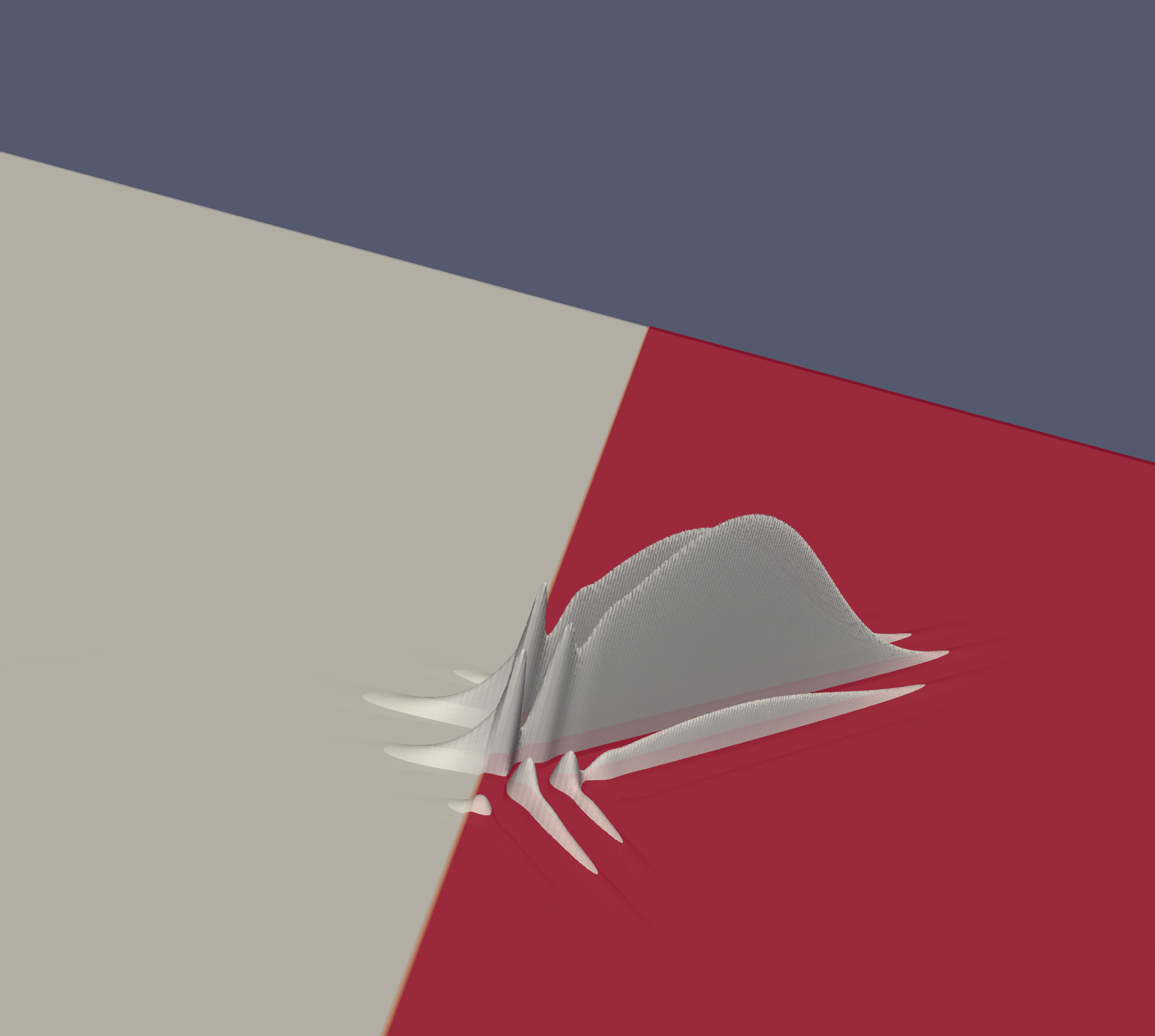}
\includegraphics[width=3.1in]{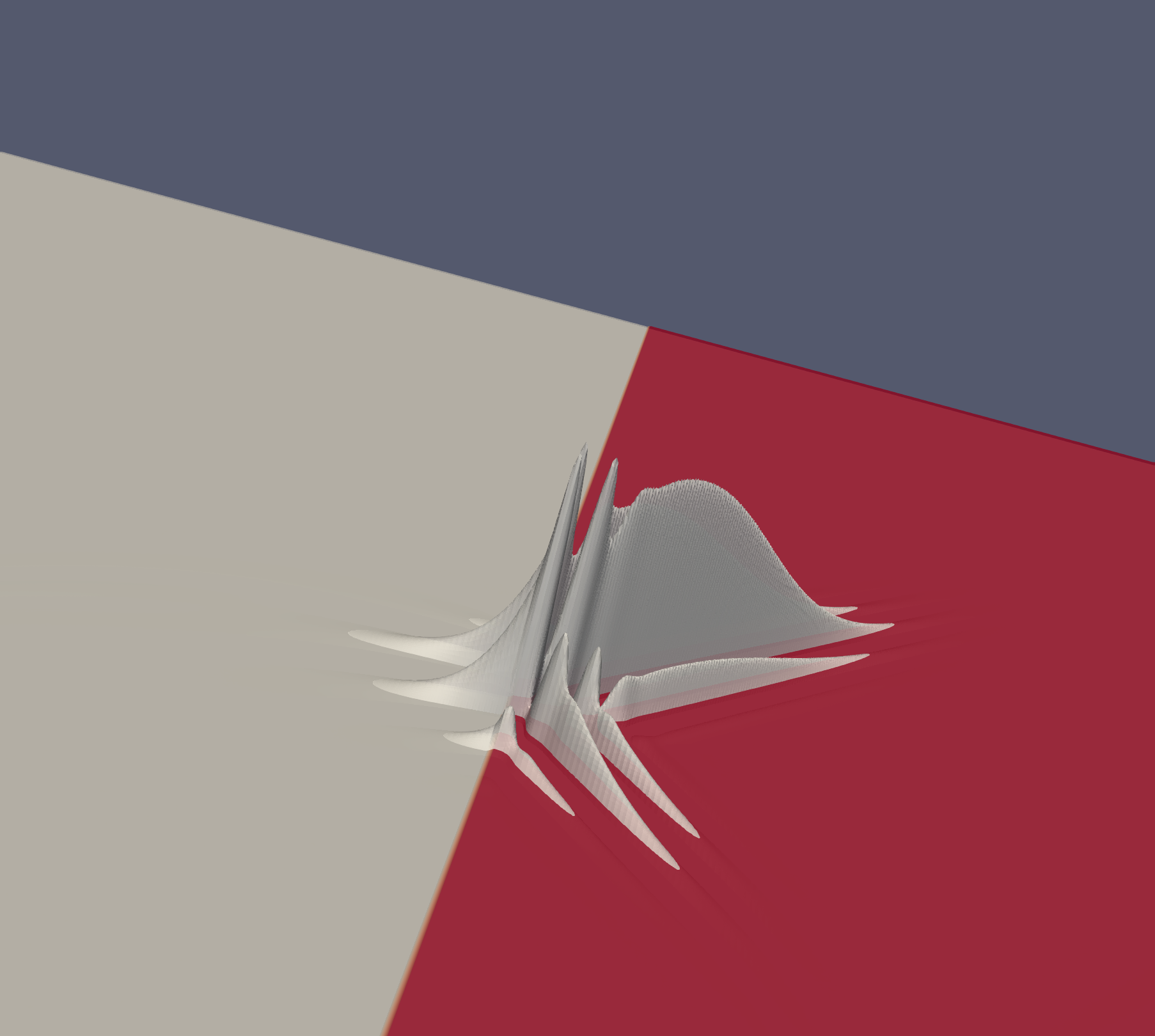}
.  (c) $H_z > 0 $ at $t = 18k$   \qquad    \qquad  \qquad \qquad (d)  $H_z > 0$ at $t = 24k$
\includegraphics[width=3.1in]{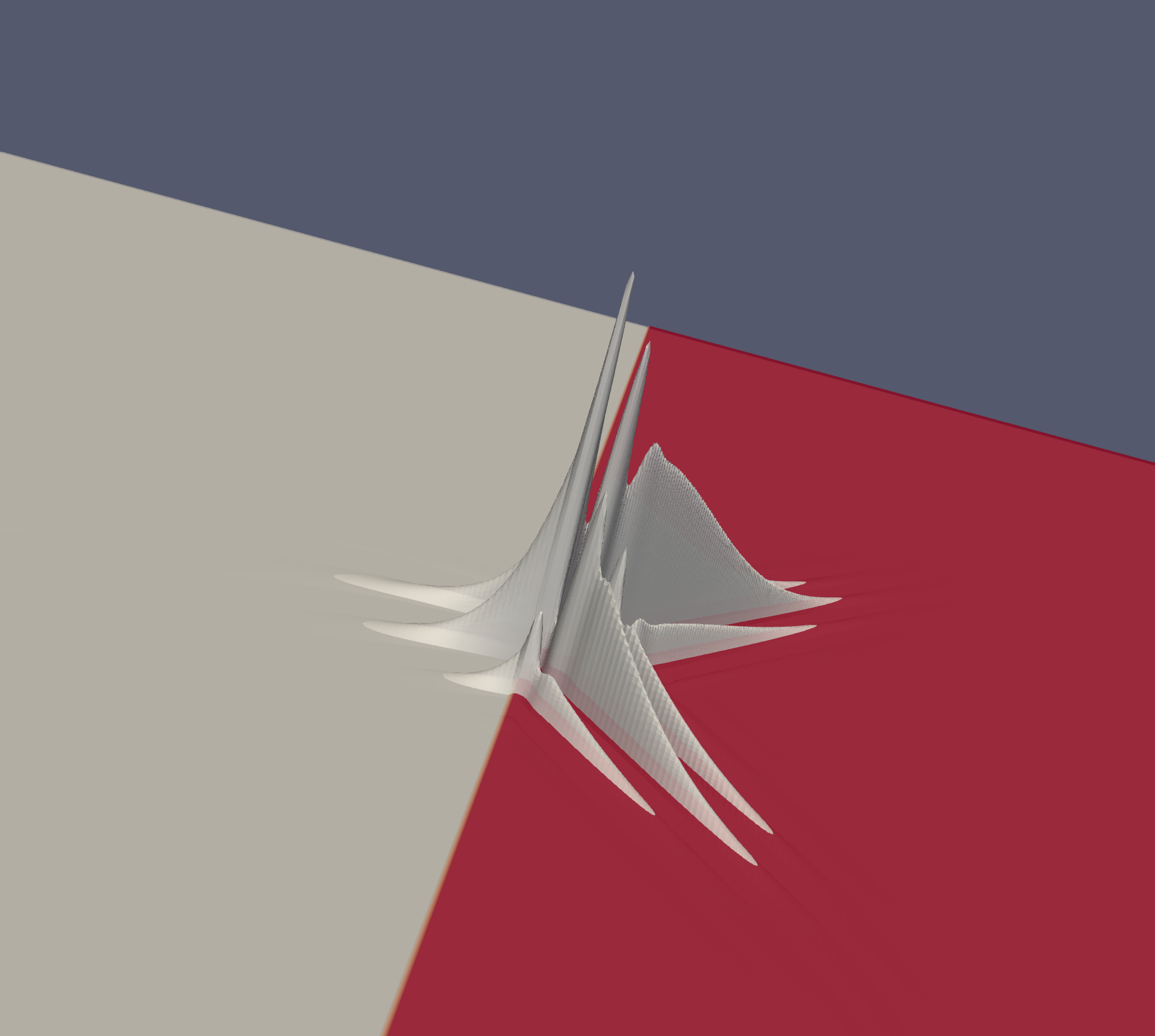}
\includegraphics[width=3.1in]{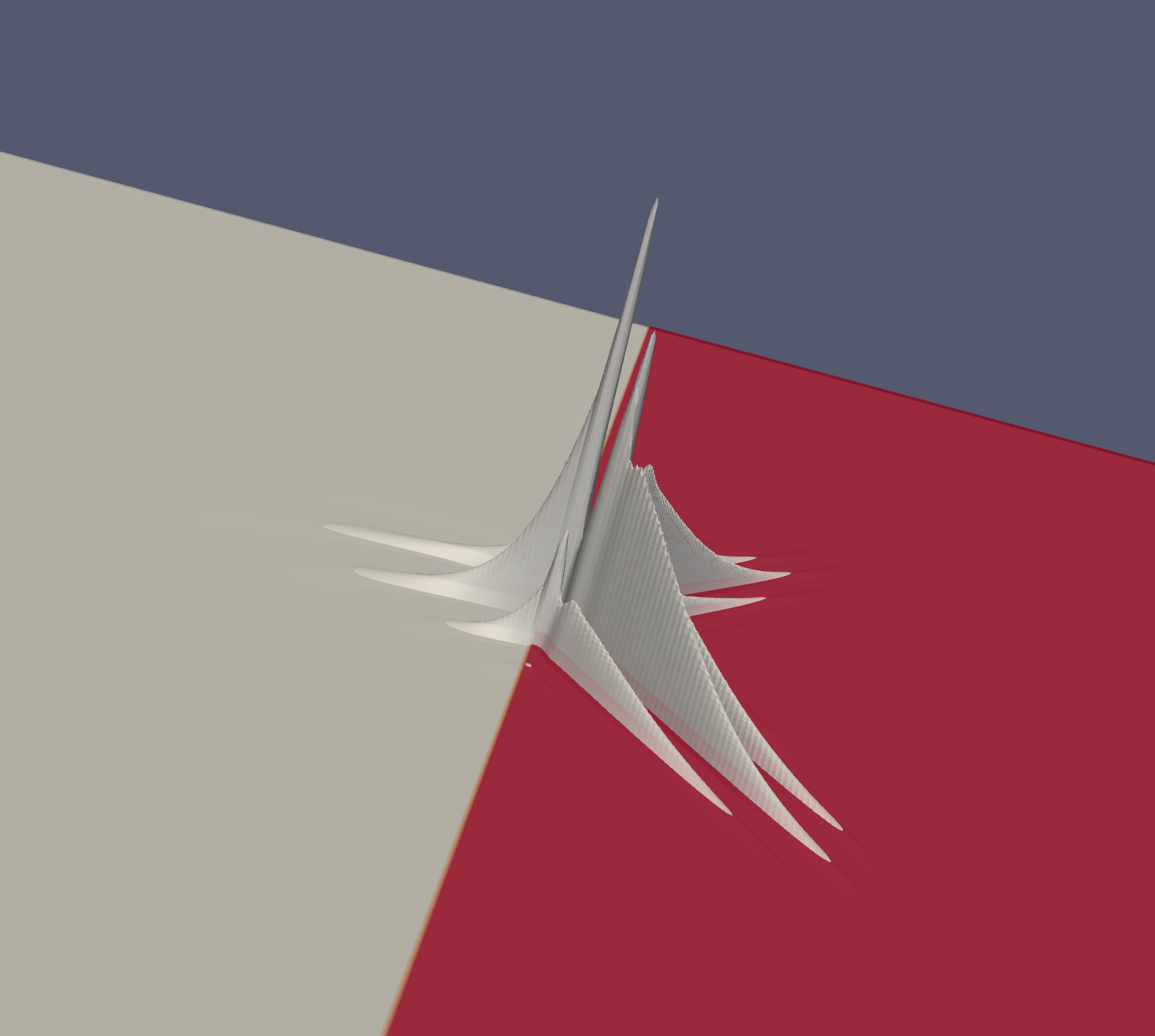}
.  (e) $H_z > 0$ at $t = 30k$   \qquad    \qquad  \qquad \qquad  (f)  $H_z > 0$ at $t_2 = 36k$
\caption{\it{Evolution of the magnetic field $H_z(x,y) > 0$  for $\theta = 35^o >\theta_c$.
Red: region $n_1=2, y < L/2$. 
   Grey: region $n_2=1 , y \ge L/2$.}
}
\end{center}
\end{figure}

 \begin{figure}[!h!p!b!t] \ 
\begin{center}
\includegraphics[width=3.2in]{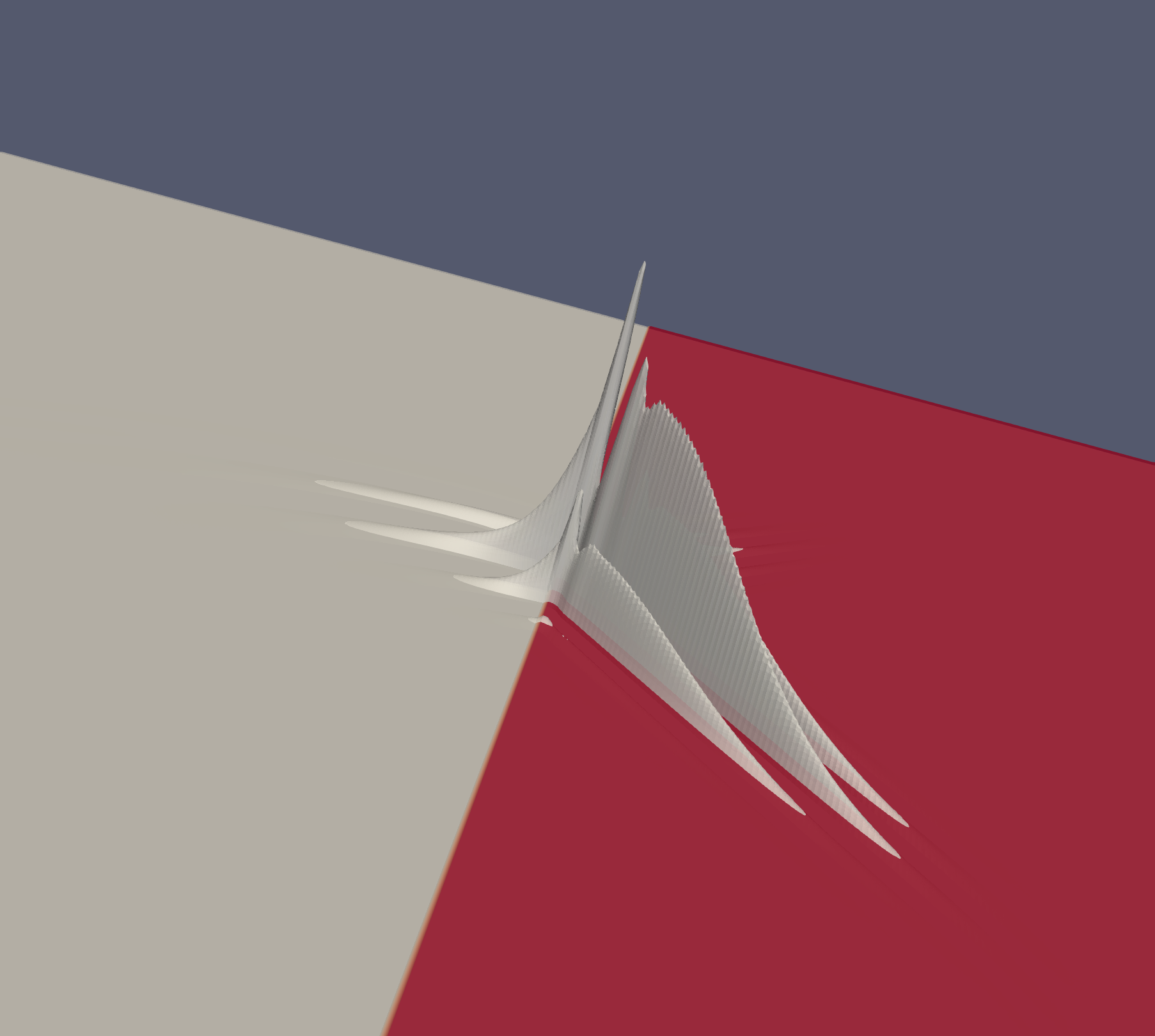}
\includegraphics[width=3.2in]{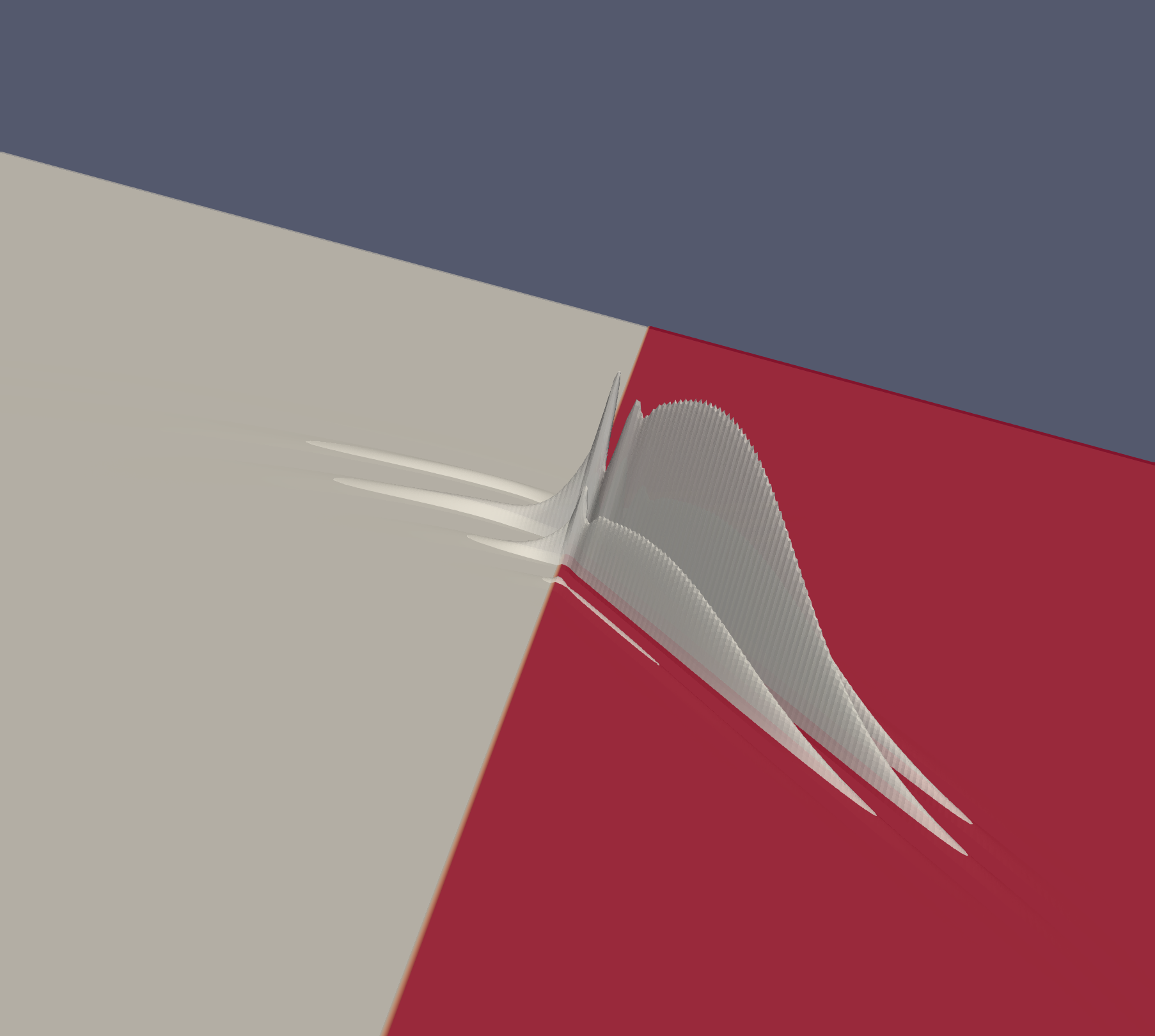}
(a) $H_z > 0 $ at $t = 42k$   \qquad ,  \qquad  \qquad  (b)  $H_z > 0$ at $t = 48k$
\includegraphics[width=3.2in]{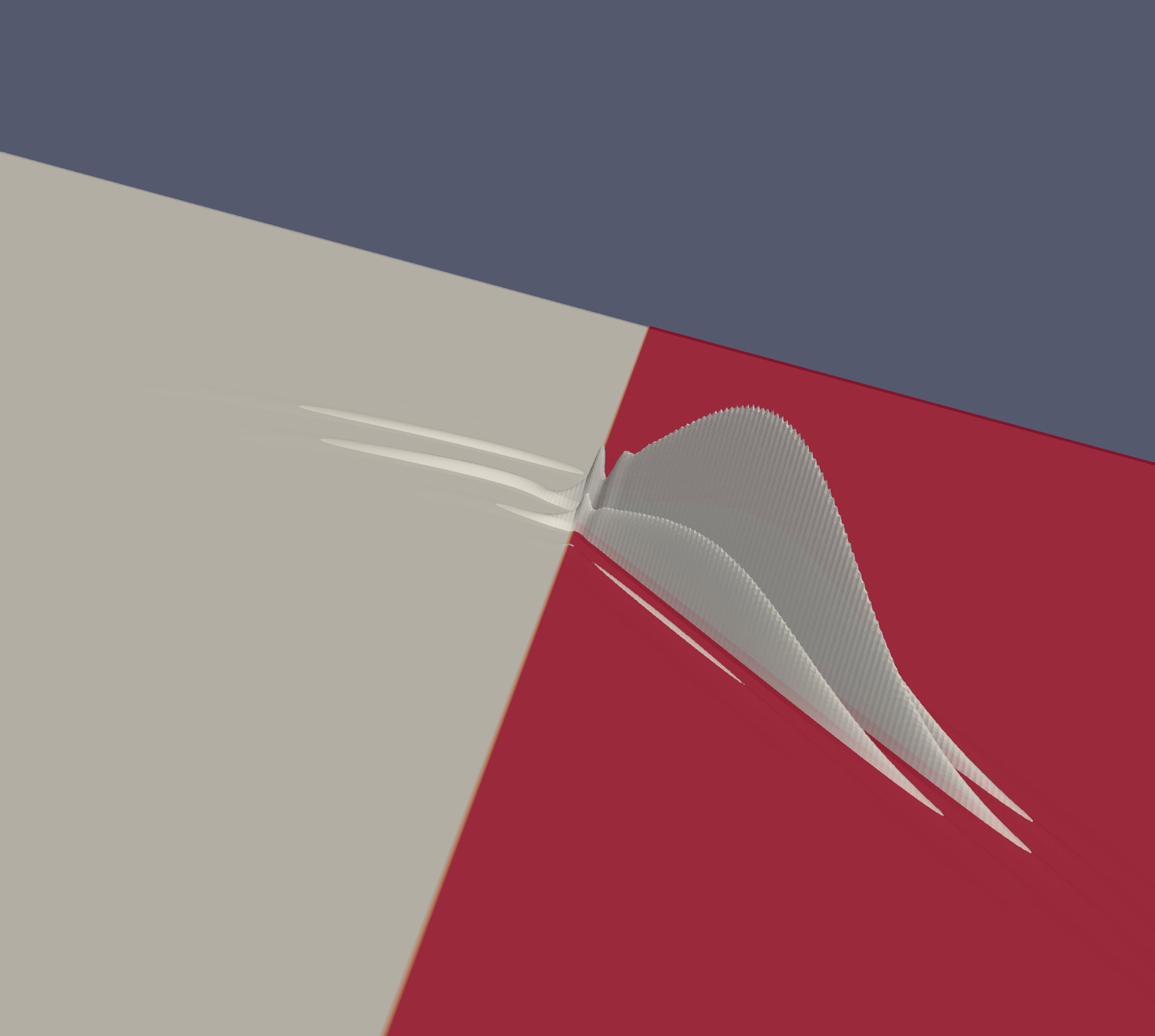}
\includegraphics[width=3.2in]{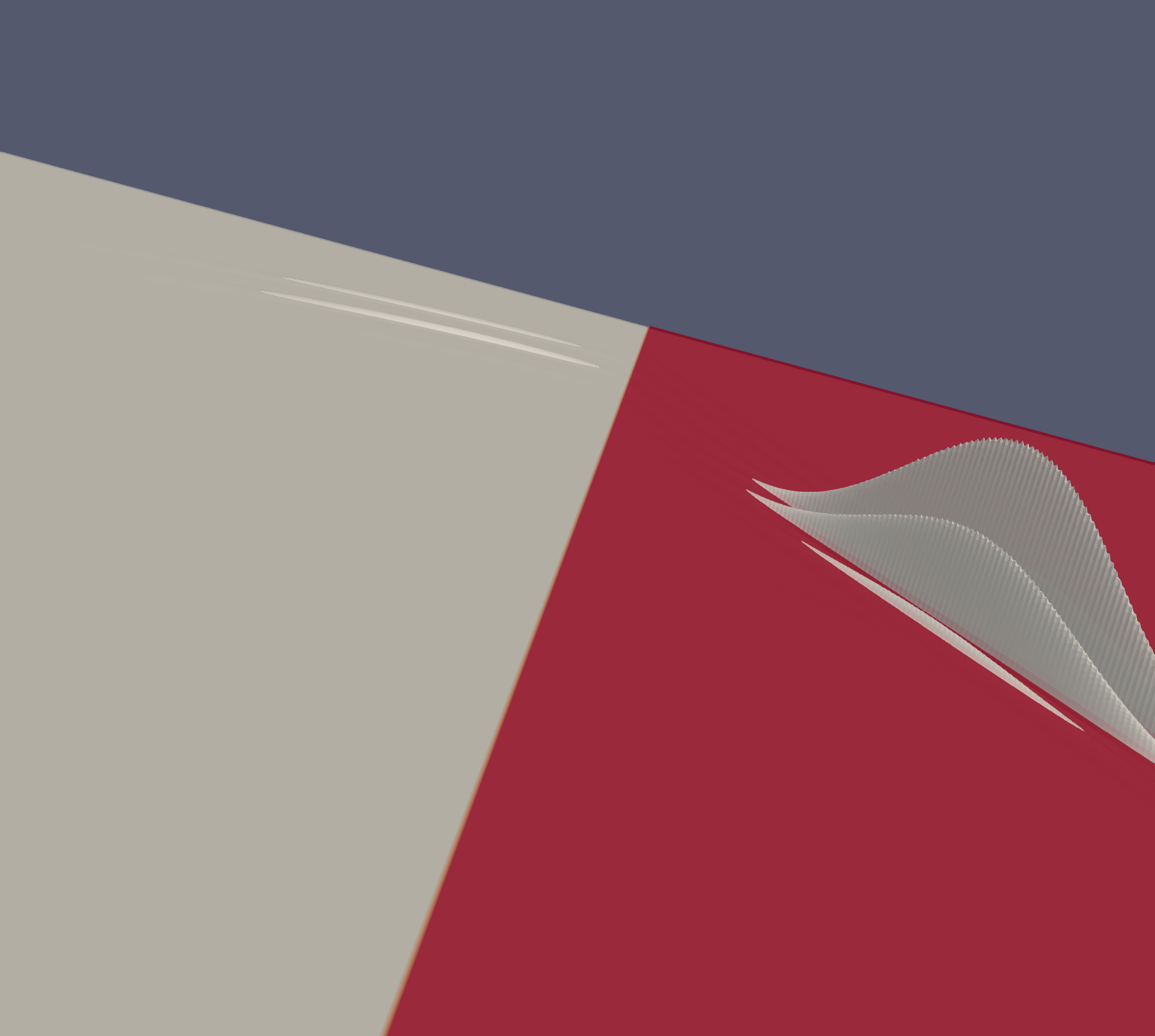}
(c) $H_z > 0 $ at $t = 54k$   \qquad ,  \qquad  \qquad  (d)  $H_z > 0$ at $t = 84k$
\caption{\it{Evolution of the magnetic field $H_z(x,y) > 0$  for $\theta = 35^o >\theta_c$ at later times.
Red: region $n_1=2, y < L/2$. 
Grey: region $n_2=1 , y \ge L/2$.  The Goos-Hanchen [9] longitudinal boundary shift is clearly seen on comparing
Fig 3b and 4c.}
}
\end{center}
\end{figure}

We also see that there is a $\pi$-like phase change in the magnetic field $H_z$ as the incident pulse in $n_1=2$ region is reflected
from the beginning of the $n_2=1$ dielectric region.  Figs 3 and 4 only plot the $H_z > 0$ surfaces, while Fig. 5 is a full 2D projection with
green for that of the profile having $H_z < 0$ and red for $H_z > 0$.  
It can be seen that the leading edge of the incident pulse has $H_z < 0$ 
while the leading edge of the reflected pulse has $H_z > 0$.
 \begin{figure}[!h!p!b!t] \ 
\begin{center}
\includegraphics[width=3.15in]{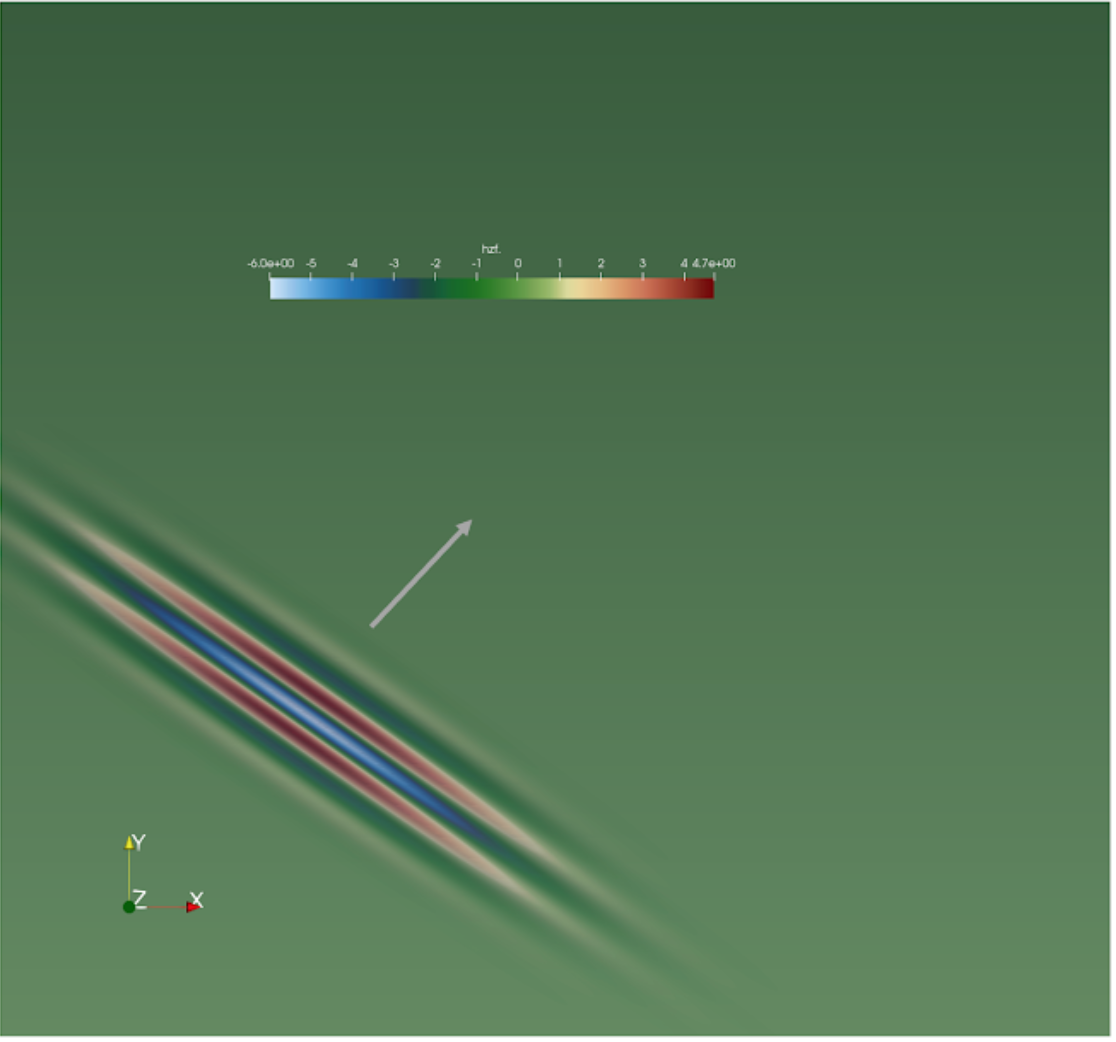}
\includegraphics[width=3.15in]{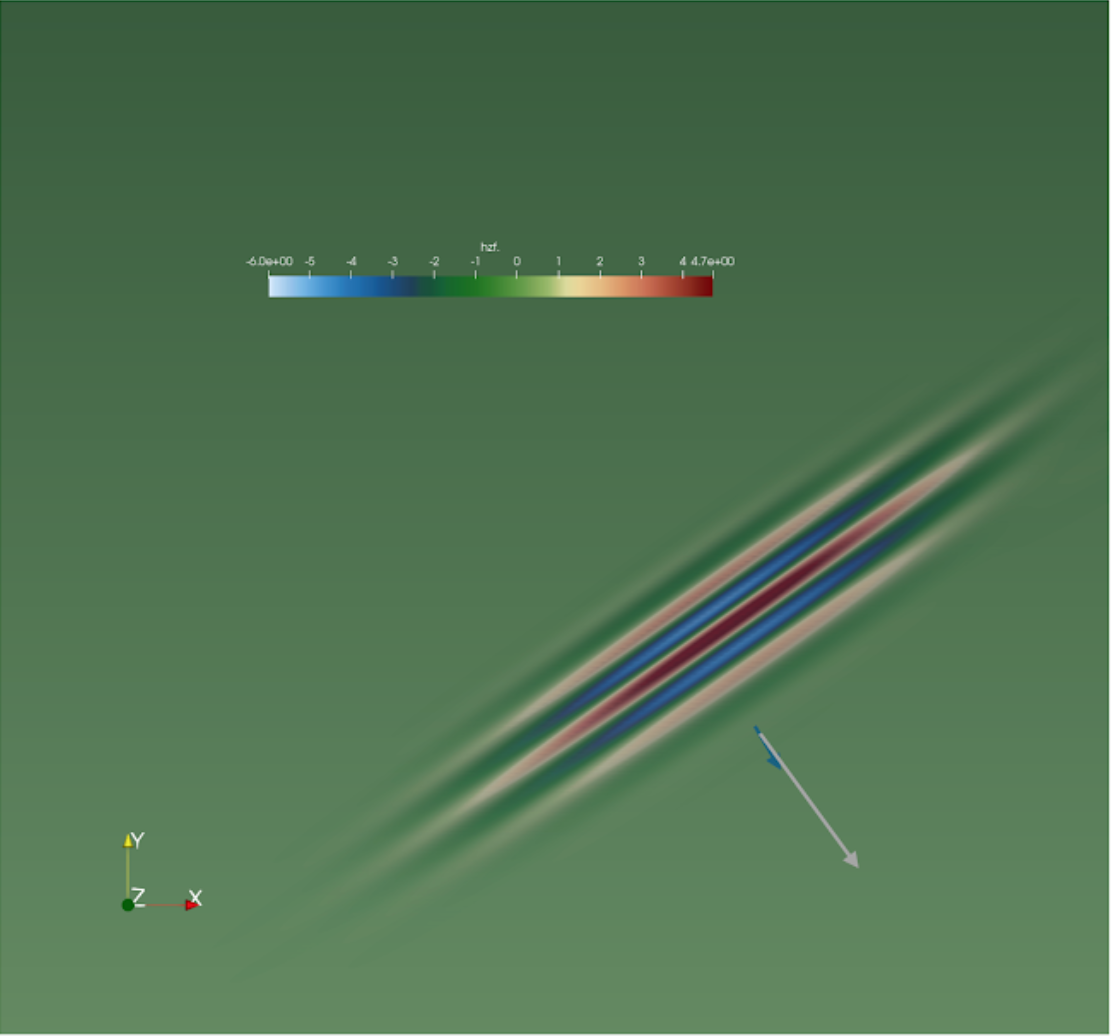}
(a) incident $H_z$ at $t=0$   \qquad ,  \qquad  \qquad  (b)  reflected $H_z $ at $t = 72k$
\caption{\it{(a) The initial magnetic field $H_z(x,y)$  at $t = 0$ as the pulse propagates to the plane dielectric boundary,
 and (b) the reflected $H_z(x,y)$ at $t = 72k$ as the pulse moves from the dielectric interface.  The color coding for the $H_z$-profile
 is shown in the horizontal strip:  blue-green for $H_z< 0$ and yellow-red for $H_z > 0$.
}
}
\end{center}
\end{figure}

\section{Fresnel Conditions for Normal Incidence}
We now consider the normal incidence of our electromagnetic pulse from medium $n_1$ onto medium $n_2$, with $n_1 < n_2$.  In Fig 6 we plot
the total energy reflected and transmitted from the $n_1 - n_2$ interface.  For these runs, $n_1 = 1$.  As $n_2 > n_1$ the 
transmitted energy decreases, while the reflected energy increases with the total energy remaining constant to 7
significant figures.
 
 \begin{figure}[!h!p!b!t] \ 
 \begin{center}
\includegraphics[width=4.5in]{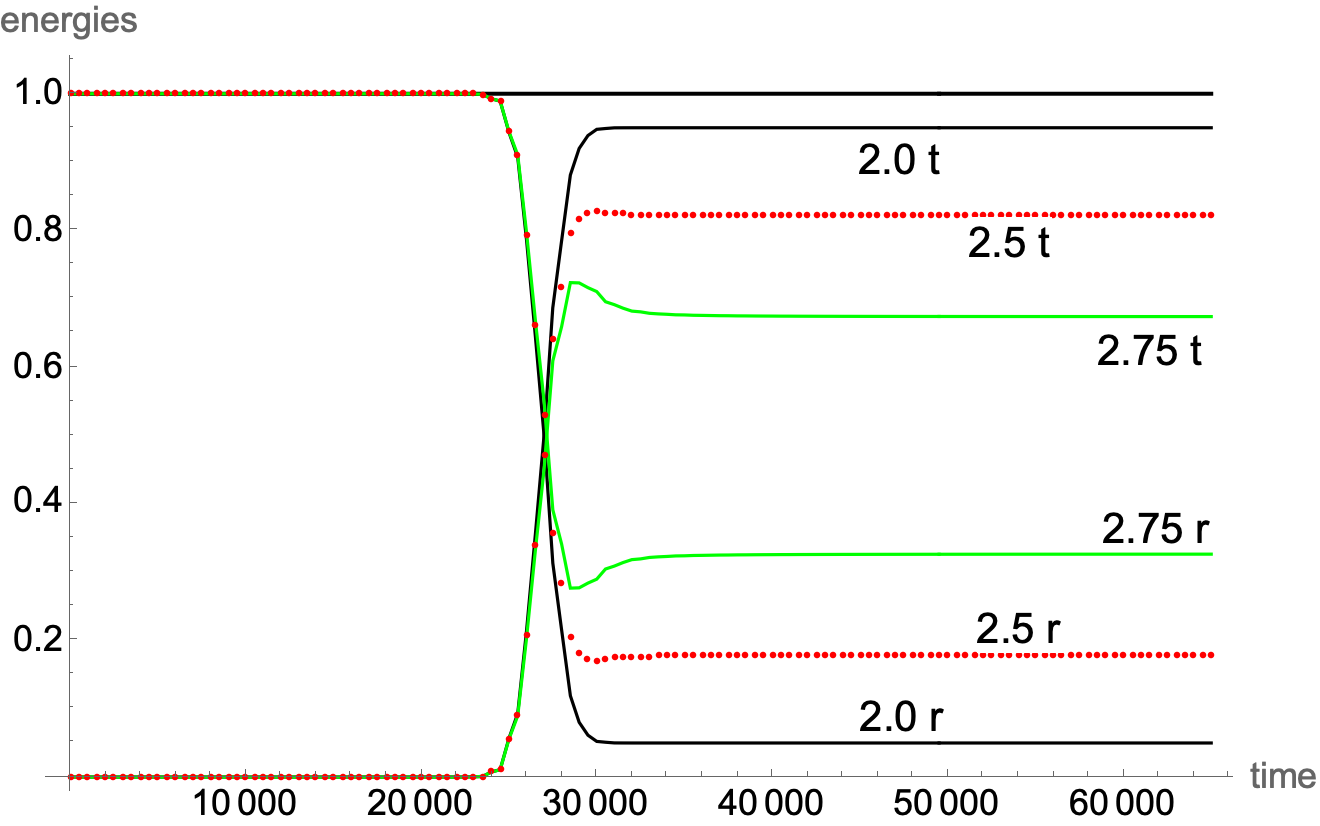}
\caption{ \it{Normal incidence.  The time evolution of the normalized reflected $"r"$ and transmitted $"t"$ energy from vacuum
$(n_1=1)$ into 3 different dielectrics $n_2 = 2.0, n_2=2.5, n_2 = 2.75$. (Nontrivial plots for $time > 25000$ when the pulse reaches
the plane interface between the two dielectric slabs.)
At every time output, the total energy
 $\mathcal{E}_1(t) + \mathcal{E}_2(t) = const.$ to the 7th significant figure.  In these simulations $L=1024$ and
energies are normalized to the initial pulse total energy $\mathcal{E}_1(0) + \mathcal{E}_2(0) = const.$  Fig 2 and Fig 6 use 
different initial pulse locations since the pulse in Fig. 2 has half the speed of that in Fig. 6.  The correlation of the time axes in
these figures is not straightforward.
}
}
\end{center}
\end{figure}

It can be readily shown that the Fresnel conditions for a simple plane wave incident, reflected and transmitted at a plane
interface, for p-polarization, satisfy [15]
\begin{equation} \label{eq3}
 r_E \equiv \frac{E_r}{E_i} = \frac{n_2 - n_1}{n_1+n_2} \quad , \quad  t_E \equiv \frac{E_t}{E_i} = \frac{2 n_1}{n_1+n_2}
 \end{equation}
 For normal incidence, the p-polarization and s-polarization are equivalent.   Thus the reflected and incident electric fields
 are in opposite directions, so that $E_r/E_t < 0$.  From Eq. (4),  $(n_2 - n_1)/(n_1+n_2) > 0$ so that the ratio  $E_r/E_i < 0$ stays negative.
  Alternatively, the reflected electric field undergoes
 a phase change of $\pi$ when hitting an interface with $n_2 > n_1$.

 For the magnetic field, in p-polarization, the incident, reflected, and transmitted fields are all in the same direction with the
 Fresnel ratio [15]
 \begin{equation} \label{eq4}
 r_H \equiv \frac{H_r}{H_i} = \frac{n_2 - n_1}{n_1+n_2} \quad , \quad  t_H \equiv \frac{H_t}{H_i} = 1+r_H
 \end{equation}
 
 The QLA calculation of the Fresnel ratio is not straightforward for our initial bounded 2D electromagnetic pulse in its reflection
 and transmission.  Since we are dealing with normal incidence all the propagation is quasi-1D :  along the $y$-axis.  We thus
 determine the position $y_m$ where the corresponding electric and magnetic fields have their maxima and then average the fields over
 their finite extent in $x$:  e,g, this average for the reflected fields is
 \begin{equation} \label{eq5}
  <E_r > = \int_0^L dx E_r(x,y_m)  \quad \, \quad   <H_r > = \int_0^L dx H_r(x,y_m)  .
 \end{equation}
  \begin{table}[ht]
 \centering
\begin{tabular}{|c|c|c||c|c|} \hline
 $\mathbf{Electric \, Field}$ &$r_E$  analytic  & $< r_E>_{QLA} \vline $ & $t_E$ analytic & $  <t_E >_{QLA} $ \\ \hline \hline
$n_2 = 1.5$ & 0.2000 & 0.2119   & 0.8000 & 0.7881  \\  \hline
$n_2 = 2.0$ & 0.3333 & 0.3507  & 0.6667 & 0.6493    \\  \hline
$n_2 = 2.5$ & 0.4286 & 0.4280  &  0.5714 & 0.5720    \\   \hline
$n_2 = 3.0$ & 0.5000 & 0.4916  & 0.5000  & 0.5084  \\  \hline  \hline
\end{tabular}
\caption{\it{The Fresnel coefficients, \eqref{eq3}, for Normal Incidence from medium $n_1=1$ to various $n_2 > n_1$ 
dielectrics for the electric field.  The QLA calculated Fresnel coefficients are derived using \eqref{eq5}.}
}
\end{table}

  \begin{table}[ht]
 \centering
\begin{tabular}{|c|c|c||c|c|} \hline
 $\mathbf{Magnetic \, Field}$ &$r_H$  analytic  & $< r_H>_{QLA} \vline $ & $t_H$ analytic & $  <t_H >_{QLA} $ \\ \hline \hline
$n_2 = 1.5$ & 0.2000 & 0.1815   & 1.200 & 1.1815  \\  \hline
$n_2 = 2.0$ & 0.3333 & 0.2975  & 1.333 & 1.298    \\  \hline
$n_2 = 2.5$ & 0.4286 & 0.4329  &  1.429 & 1.433    \\   \hline
$n_2 = 3.0$ & 0.5000 & 0.5237  & 1.500  & 1.524  \\  \hline  \hline
\end{tabular}
\caption{\it{The Fresnel coefficients, \eqref{eq3}, for Normal Incidence from medium $n_1=1$ to various $n_2 > n_1$ 
dielectrics for the magnetic field.  The QLA calculated Fresnel coefficients are derived using \eqref{eq5}.}
}
\end{table}

In Tables 1 and 2 we determine the Fresnel coefficients for normal incidence.  For the incident theoretical plane wave,
the Fresnel coefficients are immediately determined from the refractive indices, \eqref{eq3} and \eqref{eq4}, with the
reflection coefficients equal for the fields: $ r_E = r_H$.  For the QLA Fresnel coefficients, these are determined directly
from the qubit amplitude solutions and then averaging the electric and magnetic field, without explicit references to the
refractive indices themselves.

\section{Summary and Conclusion}

We have been somewhat in a quandary why QLA produces such excellent 2D and 3D scattering results for an incident 1D Gaussian
pulse onto small dielectric objects but yet 1D-QLA could not handle the relatively simple textbook problem of 1D normal incidence of a
wave packet onto  a plane dielectric slab interface.  For this 1D problem, for energy conservation to hold, the 1D-QLA requires 
the scaling of the interface region to be such that the simulations now model a 
WKB-like scattering with very little electromagnetic reflection.  
Here we come to a possible resolution of this dilemma.  Instead of restricting ourselves to only 1D, we have introduced
a 2D Gaussian wave packet that cuts off in each of these two directions.  Hence, even when restricting ourselves to 1D normal
incidence, our incident pulse is 2D and thus making the QLA simulation 2D.  We now find QLA can readily handle this 1D normal
incidence problem and from it we recover the Fresnel conditions on reflection and transmission (and preserving energy
conservation to the 7th significant figure).

In our full 2D QLA simulations we considered a p-polarized incident pulse with its $\mathbf{E}_i$ in the plane of incidence.
For this polarization the magnetic field has only one component, perpendicular to the plane of incidence.
Since QLA is an initial value code, we examined the transient rotation of the incident pulse propagation direction to that of
the reflected pulse under total internal reflection.  
Not only are there transient and spatially concentrated energy spikes in the vicinity of the interface
but there is a spatial shift in the pulse center as it hits the dielectric binterface and at which it reflects from the interface.
This is clearly related to the Goos-Hanchen effect [15] which arises because of the finte-size of the pulse 
In all the simulations reported here, a sharp boundary is assumed between the two dielectric (rather
than our usual hyperbolic-tangent buffer region).  Finally, we also find a $\pi$-like phase shift in the reflected magnetic field.  
It should be noted that in these simulations $\nabla \cdot \mathbf{H} = 0 $ since $\mathbf{H}=H_z(x,y,t) \hat{z}$ while
throughout the run $|\nabla \cdot \mathbf{D}| / D_{max}< O(10^{-19})$ .

While theoretically one can determine a unitary QLA for the energy-conserving Maxwell equations through a Dyson map [5],
it is non-trivial to explicitly determine the required unitary collision and potential operators.  Thus for the QLA presented here
to be encodable onto a quantum computer one must investigate unitary representations of potential operators like \eqref{eqA9}
in the Appendix.  

\section*{Acknowledgments}
This research was partially supported by Department of Energy grants DE-SC0021647, DE-FG02- 91ER-54109, DE-SC0021651, DE-SC0021857, and DE-SC0021653. This work has been carried out within the framework of the EUROfusion Consortium, funded by the European Union via the Euratom Research and Training Programme (Grant Agreement No. 101052200 - EUROfusion). Views and opinions expressed, however, are those of the authors only and do not necessarily reflect those of the European Union or the European Commission. Neither the European Union nor the European Commission can be held responsible for them. E. K. is supported by the Basic Research Program, NTUA, PEVE. K.H is supported by the National Program for Controlled Thermonuclear Fusion, Hellenic Republic. This research used resources of the National Energy Research Scientific Computing Center (NERSC), a U.S. Department of Energy Office of Science User Facility located at Lawrence Berkeley National Laboratory, operated under Contract No. DE-AC02-05CH11231 using NERSC award FES-ERCAP0020430.

\section*{REFERENCES}

\qquad [1]  G. Vahala, L. Vahala, M. Soe, and A. K. Ram, "Unitary quantum lattice simulations for Maxwell equations in vacuum and in dielectric media," J. Plasma Phys. 86, 905860518 (2020).

[2]  G. Vahala, L. Vahala, M. Soe, and A. K. Ram, "One- and two dimensional quantum lattice algorithms for Maxwell equations in inhomogeneous scalar dielectric media I: theory," Radiat. Eff. Defects Solids 176, 49?63 (2021). 

[3]  G. Vahala, J. Hawthorne, L. Vahala, A. K. Ram, and M. Soe, "Quantum lattice representation for the curl equations of Maxwell equations," Radiat. Eff. Defects Solids 177, 85?94 (2022). 

[4]  G. Vahala, M. Soe, L. Vahala, A. K. Ram, E. Koukoutsis, and K. Hizanidis, "Qubit lattice algorithm simulations of Maxwell?s equations for scattering from anisotropic dielectric objects," Comput. Fluids 266, 106039 (2023).

[5]  E. Koukoutsis, K. Hizanidis, A. K. Ram, and G. Vahala, "Dyson maps and unitary evolution for Maxwell equations in tensor dielectric media," Phys. Rev. A 107, 042215 (2023)

[6]  G. Vahala, M. Soe, E. Koukoutsis, K. Hizanidis, L. Vahala, and A. K. Ram, "Qubit lattice algorithms based on the Schrodinger-Dirac representation of Maxwell equations and their extensions," in 'Schrodinger Equation - Fundamentals Aspects and Potential Applications', edited by D. M. B. Tahir, D. M. Sagir, A. P. M. I. Khan, D. M. Rafique, and D. F. Bulnes (IntechOpen, Rijeka, 2023) Chap. 5.

[7]  J. Yepez, "A quantum lattice-gas model for computational fluid dynamics," Phys. Rev. E63, 046702 (2001).

[8]  J. Yepez, "Quantum lattice-gas model for Burgers equation", J. Stat. Phys. 107, 203-224 (2002)

[9]  J. Yepez, "Relativistic path integral as a lattice-based quantum algorithm", Quantum Infor. Proc. 4, 471-509 (2005)

[10]  M. A. Pravia, Z. Chen, J. Yepez and D. G. Cory, "Experimental Demonstration of Quantum Lattice Gas Computation", arXiv:0303183
(2003)

[11]  J. Yepez, "Quantum lattice gas model of Dirac particles in 1+1 dimensions", arXiv:1307.3595 (2013)

[12]  E. Koukoutsis, K. Hizanidis, G. Vahala, M. Soe, L. Vahala, and A. K. Ram, "Quantum computing perspective
for electromagnetic wave propagation in cold magnetized plasmas", Phys. Plasmas 30, 122108 (2023).

[13] E. Koukoutsis, K. Hizanidis, A. K. Ram, and G. Vahala, "Quantum simulation of dissipation for Maxwell equations
in dispersive media", Future Gener. Comput. Syst. 159, 221 (2024).

[14]  R. H. Renard, "Total reflection: A new evaluation of the Goss-Hanchen shift", J, Optical Soc. America 54, 1190-1197 (1964);
F. Goos and H. Hanchen, Ann. Phys. (Leipzig) 436, 333 (1947).

[15]  A. Zangwill, "Modern Electrodynamics", Cambridge Univ Press, 2012.

[16] A. M. Childs and N. Wiebe, "Hamiltonian simulation using linear combinations of unitary operations", Quantum Info. Comput. 12, 901?924 (2012).

[17]  A. Mostafazadeh, "Pseudo-Hermitian Representation of Quantum Mechanics", Int. J. Geom. Meth. Mod. Phys. 07, 1191 (2010)

[18]  M. Znojil, Quantum mechanics using two auxiliary inner products, Phys. Lett. A 421, 127792 (2022).

[19] A. Fring and M. H. Y. Moussa, Unitary quantum evolution for time-dependent quasi-Hermitian systems with nonobservable Hamiltonians, Phys. Rev. A 93, 042114 (2016).

[20]  E. Esmaeilifar, D. Ahn, R. S. Myong, "Quantum algorithm for nonlinear Burgers' equation for high-speed compressible flows",
Phys. Fluids 36, 106110 (2024)

\section*{Appendix [6]}
The time evolving subset of the Maxwell equations for non-magnetic inhomogeneous dielectric media is
$\partial \mathbf{B}/ \partial t=-\nabla \times \mathbf{E} $, and $\partial \mathbf{D} / \partial t=\nabla \times \mathbf{H}$, with $\mathbf{D} =  \mathcal{\epsilon}\cdot \mathbf{E}$ and  $\mathbf{B}=\mu_{0} \mathbf{H}$ .  The divergence subset of Maxwell equations can be treated as initial conditions.  If one works with the field
$\mathbf{u}=(\mathbf{E}, \mathbf{H})^{\mathbf{T}}$ then the matrix Maxwell equations representation
\begin{equation} \label{eqA1}
i \frac{\partial \mathbf{u}}{\partial t}=\mathbf{W}^{-\mathbf{1}} \mathbf{M} \mathbf{u}
\end{equation}
is non-unitary for inhomogeneous dielectric media since $\mathbf{W}^{-1}$ and $\mathbf{M}$ will no longer commute.  
Here $\mathbf{W}$ is a Hermitian $6 \times 6$ matrix for lossless media and $\mathbf{M}$ the Hermitian
curl-matrix operator under standard boundary conditions:
\begin{equation} \label{eqA2}
\mathbf{W}=\left[\begin{array}{cc}
\epsilon_i \mathbf{I}_{3 \times 3} & 0_{3 \times 3} \\
0_{3 \times 3}& \mu_{0} \mathbf{I}_{3 \times 3}
\end{array}\right]   \, , \quad  \mathbf{M}=\left[\begin{array}{cc}
0_{3 \times 3} & i \nabla \times \\
-i \nabla \times & 0_{3 \times 3}
\end{array}\right]  .
\end{equation}
where we work in a coordinate system where the Hermitian dielectric tensor is diagonal $\epsilon_i , i = x,y,z$.

There exists a Dyson map [5], [10]-[12]
\begin{equation} \label{eqA3}
\mathcal{U} = \mathbf{W}^{1/2} \mathbf{u}
\end{equation}
that yields a unitary evolution for the $\mathcal{U}$-field even for inhomoegneous dielectric media:
\begin{equation} \label{eqA4}
i \frac{\partial \mathcal{U}}{\partial t}=\mathbf{W}^{-1 / 2} \mathbf{M} \mathbf{W}^{-1 / 2} \mathcal{U}
\end{equation}
\noindent since $\mathbf{W}^{-\mathbf{1} / \mathbf{2}} \mathbf{M} \mathbf{W}^{-\mathbf{1} / \mathbf{2}}$ is Hermitian.
For non-magnetic materials, \eqref{eqA3} is 
\begin{equation} \label{eqA5}
\mathcal{U}=\left(\epsilon_0^{1/2} n_x E_x, \epsilon_0^{1/2}  n_y E_y,  \epsilon_0^{1/2} n_z E_z, \mu_0^{1/2}  \mathbf{H}\right)^{T}    .
\end{equation}
where $(n_x , n_y, n_z)$ is the vector (diagonal) refractive index, with $\epsilon_i =\epsilon_0  n_i^2, i=x,y,z$ ,
and \eqref{eqA4} for 2D $x-y$ spatially dependent fields is
\begin{equation} \label{eqA6}
\begin{aligned}
\frac{\partial q_0}{\partial t} = \frac{1}{n_x} \frac{\partial q_5}{\partial y} , \qquad
\frac{\partial q_1}{\partial t} = - \frac{1}{n_y} \frac{\partial q_5}{\partial x} , \qquad
\frac{\partial q_2}{\partial t} =  \frac{1}{n_z} \left[ \frac{\partial q_4}{\partial x} -\frac{\partial q_3}{\partial y} \right] \\
\frac{\partial q_3}{\partial t} = - \frac{\partial (q_2/n_z)}{\partial y} , \qquad
\frac{\partial q_4}{\partial t} = \frac{\partial (q_2/n_z)}{\partial x} , \qquad
\frac{\partial q_5}{\partial t} = - \frac{\partial (q_1/n_y)}{\partial x}  + \frac{\partial (q_0/n_x)}{\partial y}   ,
\end{aligned}
\end{equation}
with $\mathcal{U} \equiv (q_0, q_1, q_2, q_3, q_4, q_5)^T$.

\subsection*{Qubit Lattice Algorithm (QLA) for Eq. \eqref{eqA6}}
In moving to a qubit amplitude discrete representation, at this stage, one will need only 3 qubits per lattice site.
For the $6$ qubit amplitudes $(q_0, q_1, q_2, q_3, q_4, q_5)$. QLA consists of an appropriately chosen
 sequence of interleaved collision and streaming operators,
where the collision operators act  only locally at each lattice site while the streaming operators move these
entangled amplltudes to neighboring lattice sites.  QLA is modular in that each directional derivative can be
handled independently.  QLA is also perturbative with the lattice step size $\delta$ as perturbation parameter.

We now outline how to recover the $\partial / \partial x$-derivatives in the amplitudes in \eqref{eqA6}.  One sees coupling between
the amplitudes $q_1 - q_5$ and between the amplitudes $q_2 - q_4$.  We thus choose the unitary collision
matrix $\hat{C}_X$ to have the form
\begin{equation} \label{eqA7}
C_X=\left[\begin{array}{cccccc}
1 & 0 & 0 & 0& 0& 0 \\
0 & cos \,\theta_1 & 0 & 0 & 0 & - sin\,\theta_1 \\
0 & 0 & cos\,  \theta_2 & 0 & - sin \,\theta_2 & 0 \\
0 & 0 & 0 & 1 &  0 & 0  \\
0 & 0 & sin\,\theta_2 & 0 & cos\, \theta_2 & 0 \\
0 & sin\, \theta_1 & 0 & 0 & 0 & cos \,\theta_1
\end{array}\right]
\end{equation}
where the collision angles $\theta_1, \theta_2$ in these couplings are different for medium refractive indices $n_y \ne n_z$.
We now stream 2 amplitudes along the $x$-axis, while keeping the other 4 amplitudes fixed.  The appropriate streamed amplitudes
are $q_1 - q_4$ and the other set $q_2 - q_5$ thus coupling the two post-collision pairs in $\hat{C}_X$.
In particular, we define $S_{14}^{x+}$ to stream the amplitudes $q_1$ and $q_4$  one lattice unit in the positive $x$-direction while keeping the remaining amplitudes $q_0, q_2, q_3, q_5$ unstreamed.  Streaming is simply a shift operator and so is unitary.
The final unitary collide-stream sequence for the $x$-direction is
\begin{equation} \label{eqA8}
\mathbf{U_X} = S^{x+}_{25}.C_X^\dag . S^{x-}_{25}.C_X. S^{x-}_{14}.C_X^\dag . S^{x+}_{14}.C_X .S^{x-}_{25}.C_X . S^{x+}_{25}.C_X^\dag. S^{x+}_{14}.C_X . S^{x-}_{14}.C_X^\dag 
\end{equation}.
where $C_X^\dag$ is the adjoint of $C_X$.

Similarly to recover the $\partial / \partial y$ derivatives on the amplitudes in \eqref{eqA6}.  One sees that $q_0 - q_5$ are
coupled, as are $q_2 - q_3$.  The corresponding collision matrix $C_Y$ will introduce a new collision angle $\theta_0$ if 
dealing with a biaxial medium.  The unitary streaming operator in the $y$-direction will stream the couplet $q_0 - q_3$ and
the couplet $q_2 - q_5$ with the other qubits unstreamed.

However, we still need to recover terms that involve spatial derivatives on the refractive indices.  In particular to recover the
$\partial n_y / \partial x$ term we need to couple amplitudes $q_5 - q_1$, while to recover $\partial n_z / \partial x$ we couple
$q_4 - q_2$.  The resulting potential matrix is non-unitary with
\begin{equation} \label{eqA9}
V_X=\left[\begin{array}{cccccc}
1 & 0 & 0 & 0& 0& 0 \\
0 & 1 & 0 & 0 & 0 & 0\\
0 & 0 & 1 & 0 &0& 0 \\
0 & 0 & 0 & 1 &  0 & 0  \\
0 & 0 &- sin\,\beta_2 & 0 & cos\, \beta_2 & 0 \\
0 & sin\, \beta_0 & 0 & 0 & 0 & cos \,\beta_0
\end{array}\right]
\end{equation}
Similarly for the corresponding non-unitary $V_Y$ for the recovery of  $\partial n_x / \partial y$ and $\partial n_z / \partial y$.
Thus the QLA time stepping initial value code evolves as
\begin{equation} \label{QLA_time_step}
   \mathcal{U}(t+1) = V_Y V_X \mathbf{U_Y U_X} \, \mathcal{U}(t).
\end{equation}
   
Note that one can readily find a weighted sum of unitary matrices to obtain $V_X$.  First consider the simple unitary matrix 
from the structure of the matrix $V_X$ :
\begin{equation} \label{eqA9a}
U0_X=\left[\begin{array}{cccccc}
1 & 0 & 0 & 0& 0& 0 \\
0 & cox[\beta_0] & 0 & 0 & 0 & -sin[\beta_0]\\
0 & 0 & cos[\beta_2] & 0 &sin[\beta_2]& 0 \\
0 & 0 & 0 & 1 &  0 & 0  \\
0 & 0 &- sin\,\beta_2 & 0 & cos\, \beta_2 & 0 \\
0 & sin\, \beta_0 & 0 & 0 & 0 & cos \,\beta_0
\end{array}\right]
\end{equation}
Another unitary matrix constructed from the structure of $V_X$, but which would eliminate the new off-diagonal elements
in $U0_X$ is $U1_X$:
\begin{equation} \label{eqA9b}
U1_X=\left[\begin{array}{cccccc}
1 & 0 & 0 & 0& 0& 0 \\
0 & -cox[\beta_0] & 0 & 0 & 0 & sin[\beta_0]\\
0 & 0 & -cos[\beta_2] & 0 &-sin[\beta_2]& 0 \\
0 & 0 & 0 & 1 &  0 & 0  \\
0 & 0 &- sin\,\beta_2 & 0 & cos\, \beta_2 & 0 \\
0 & sin\, \beta_0 & 0 & 0 & 0 & cos \,\beta_0
\end{array}\right]
\end{equation}
so that the weighted sum of the unitary matrices $(U0_X+U1_X)/2$ yields
\begin{equation} \label{eqA9c}
\frac{1}{2} (U0_X + U1_X) =\left[\begin{array}{cccccc}
1 & 0 & 0 & 0& 0& 0 \\
0 & 0 & 0 & 0 & 0 &0 \\
0 & 0 & 0 & 0 & 0 & 0 \\
0 & 0 & 0 & 1 &  0 & 0  \\
0 & 0 &- sin\,\beta_2 & 0 & cos\, \beta_2 & 0 \\
0 & sin\, \beta_0 & 0 & 0 & 0 & cos \,\beta_0  
\end{array}\right] .
\end{equation}
On comparing \eqref{eqA9c} with \eqref{eqA9}, one needs to find a weighted sum of unitary diagonal  matrices to recover just 
the missing diagonal elements in \eqref{eqA9c}.   These two unitary diagonal matrices are simply the identity matrix
 $U2_X$ and $U3_X$:
\begin{equation} \label{eqA9d}
U2_X=\left[\begin{array}{cccccc}
1 & 0 & 0 & 0& 0& 0 \\
0 & 1 & 0 & 0 & 0 &0 \\
0 & 0 & 1 & 0 & 0 & 0 \\
0 & 0 & 0 & 1 &  0 & 0  \\
0 & 0 & 0 & 0 & 1 & 0 \\
0 & 0 & 0 & 0 & 0 & 1  
\end{array}\right]  , \qquad   \text{and} \qquad
U3_X=\left[\begin{array}{cccccc}
-1 & 0 & 0 & 0& 0& 0 \\
0 & 1 & 0 & 0 & 0 &0 \\
0 & 0 & 1 & 0 & 0 & 0 \\
0 & 0 & 0 & -1 &  0 & 0  \\
0 & 0 & 0 & 0 & -1 & 0 \\
0 & 0 & 0 & 0 & 0 & -1  
\end{array}\right] .
\end{equation}
Thus the LCU's to recover the potential operator $V_X$, \eqref{eqA9}, is
\begin{equation}
V_X = \frac{1}{2} \left [U0_X+U1_X + U2_x + U3_X \right].
\end{equation}
One finds a similar LCU decomposition of the extneral potential operator $V_Y$.  Much work has been done in quantum computing
in representing LCUs [16]-[19].

The LCU decomposition process is not unique.  For example, 
one could analytically determine the Singular Value Decomposition of $V_X$ into a product of
3 matrices, $V_X = UU.DD.VV$ where $UU$ and $VV$ are unitary and $DD$ is non-unitary but diagonal.  
By renormalizing the $DD$-matrix by its maximal eigenvalue the resulting diagonal elements will have a maximal value of 1.
This resulting matrix can be immediately split into the sum of 2 unitary operators. 

The required 2D Maxwell equations \eqref{eqA6} are recovered from our QLA sequence of operators provided the collision
and external operator matrices are chosen:
\begin{equation} \label{eqA10}
\theta_0 = \frac{\delta}{4 n_x} \quad , \qquad \theta_1 = \frac{\delta}{4 n_y} \quad , \qquad  \theta_2 = \frac{\delta}{4 n_z} ,
\end{equation}
\noindent and
\begin{equation} \label{eqA11}
\beta_0 = \delta^2 \frac{\partial n_y/\partial x}{n^2_y} \quad , \quad \beta_1 = \delta^2 \frac{\partial n_x/\partial y}{n^2_x}  \quad , \quad \beta_2 = \delta^2 \frac{\partial n_z/\partial x}{n^2_z} \quad , \quad \beta_3 = \delta^2 \frac{\partial n_z/\partial y}{n^2_z}
\end{equation}
on using symbolic manipulations.  Note the ordering of these angles in the perturbation parameter $\delta$.  Indeed, on passing 
to the continuum limit we recover the following equations
\begin{equation} \label{eqA12}
\begin{aligned}
\frac{\partial q_0}{\partial t} =\frac{ \delta^2}{ \Delta t} \frac{1}{n_x} \frac{\partial q_5}{\partial y} + O(\frac{\delta^4}{\Delta t})\\
\frac{\partial q_1}{\partial t} = -\frac{ \delta^2}{  \Delta t}  \frac{1}{n_y} \frac{\partial q_5}{\partial x} + O(\frac{\delta^4}{\Delta t})\\
\frac{\partial q_2}{\partial t} =\frac{  \delta^2}{ \Delta t}   \frac{1}{n_z} \left[ \frac{\partial q_4}{\partial x} -\frac{\partial q_3}{\partial y} \right] + O(\frac{\delta^4}{\Delta t})\\
\frac{\partial q_3}{\partial t} = -\frac{ \delta^2}{ \Delta t}  \left[ \frac{1}{n_z} \frac{\partial q_2}{\partial y}  - \frac{\partial n_z/\partial y}{n_z^2} q_2 \right]+ O(\frac{\delta^4}{\Delta t})\\
\frac{\partial q_4}{\partial t} = \frac{ \delta^2}{ \Delta t}  \left[ \frac{1}{n_z} \frac{\partial q_2}{\partial x}  - \frac{\partial n_z/\partial x}{n_z^2} q_2 \right]+ O(\frac{\delta^4}{\Delta t})\\
\frac{\partial q_5}{\partial t} = \frac{\delta^2}{ \Delta t } \left[ -\frac{1}{n_y} \frac{\partial q_1}{\partial x}  +\frac{\partial n_y/\partial x}{n_y^2} q_1 + \frac{1}{n_x} \frac{\partial q_0}{\partial y}  -\frac{\partial n_x/\partial y}{n_x^2} q_0  \right] + O(\frac{\delta^4}{\Delta t})\
\end{aligned}
\end{equation}
which recover \eqref{eqA6} under diffusion ordering ($\Delta t \approx \delta^2$) to second order in $\delta$.

It is a nontrivial task to determine a QLA that is fully unitary.  A non-unitary QLA  creates no obstacle for its extreme parallelization on 
classical supercomputers.

\subsection*{Conservation of Instantaneous Total Electromagnetic Energy }
The total electromagnetic energy  $\mathcal{E}(t)$ for our Maxwell equations
is a constant of the motion.  In Dyson variables, $\mathcal{U}$, \eqref{eqA5},
\begin{equation} \label{eqA13}
\mathcal{E}(t) =  \int_0^L \int_0^L dx dy \left[ \epsilon_0( n_x^2 E_x^2 + n_y^2 E_y^2 + n_z^2 E_z^2 )+  \mu_0 \mathbf{H}^2 \right] 
= ||\, \mathcal{U}||^2 = const.
\end{equation}  
where the total system is in a square box of length $L$.
However, since our current QLA is not fully unitary, this energy must be monitored in our simulations $\mathcal{E}_{QLA}(t)$
since it  is not a constant of the motion.  We find in all the simulations reported in Sec. 2 and 3, the energy $\mathcal{E}_{QLA}(t)$ was indeed
a constant. 

\subsection*{Unitary representation of LCUs [20]}
The state of a qubit is a linear combination of two orthogonal states $\ket{0}$ and $\ket{1}$.  In matrix representation
\begin{equation} \label{eqA14}
\ket{0} = \begin{bmatrix}
1 \\
0
\end{bmatrix} \quad , \quad \ket{1} = \begin{bmatrix}
0  \\
1
\end{bmatrix} .
\end{equation}
Here we shall discuss how to determine a unitary algorithm to determine the sum of qubit states $\ket{q_0} - a \ket{q_1}$.
This can then be generalized to handle LCUs.   Let unitary operators  $U$ and $V$ be such that on the $\ket{0}$
\begin{equation} \label{eqA15}
U \ket{0} = \ket{q_0}   \quad  ,  \quad  V \ket{0} = \ket{q_1}  .
\end{equation}
Consider the initial state of the 2-qubit system to be $\ket{0 \, 0} = \ket{0} \otimes \ket{0}$. where $\otimes$ is the tensor product.
Now apply the rotation matrix $R(\phi)$ to the first qubit (reading from right to left of $\ket{0 \, 0}$)
\begin{equation} \label{eqA16}
R(\phi) \ket{0} = \begin{bmatrix}
cos\, \phi & - sin\, \phi  \\
sin\, \phi & cos \, \phi 
\end{bmatrix} \begin{bmatrix}
1  \\
0
\end{bmatrix}  = \begin{bmatrix}
cos\, \phi   \\
sin\,  \phi   
\end{bmatrix}  = cos\, \phi \ket{0} +  sin\, \phi \ket{1}
\end{equation}
so that
\begin{equation} \label{eqA17}
R(\phi) \ket{0 \, 0} = \ket{0}  \otimes (cos\, \phi \ket{0} +  sin\, \phi \ket{1})
\end{equation}
We now apply the $U$-matrix on the second qubit $\ket{0}$ provided the first qubit is $\ket{0}$, and
$V$-matrix also on the second qubit $\ket{0}$ provided the first qubit is $\ket{1}$.  Thus
\begin{equation} \label{eqA18}
U \, V \, R(\phi) \ket{0 \, 0}  = cos\, \phi \ket{q_0 \, 0}  +  sin\, \phi \ket{q_1 \, 1}   .
\end{equation}
Now apply the Hadamard gate, H,  on the first qubit of \eqref{eqA18}, where
\begin{equation} \label{eqA19}
 H \ket{0} = \frac{1}{\sqrt 2} \left( \ket{0} + \ket{1} \right) \quad , \quad  H \ket{1} = \frac{1}{\sqrt 2} \left( \ket{0} - \ket{1} \right) 
 \end{equation}
 so that with a little rearranging one obtains
 \begin{equation} \label{eqA20}
 H \, U \, V \, R(\phi) \ket{0 \, 0}  = \frac{cos \, \phi \ket{q_0} + sin \, \phi \ket{q_1}}{\sqrt 2} \otimes \ket{0} +
  cos \, \phi \, \frac{ \ket{q_0} - tan \, \phi \ket{q_1}}{\sqrt 2} \otimes \ket{1}.
 \end{equation}
 
 Thus, applying the 4 unitary operators on the $\ket{0 \, 0}$ we obtain for the second qubit the desired state $\ket{q_0} - a \ket{q_1}$,
 provided the rotation angle $\phi$ is such that $tan \, \phi = a$.  We can uncouple this desired state from \eqref{eqA20} by applying the
 non-unitary projection operator $P = \ket{1} \bra{1}$ on the first qubit of  \eqref{eqA20} since $\bra{1} \ket{0} = 0$.  Extending this to the
 linear combination of unitaries, we can determine a unitary set of operators which will, with some probability, recovery the desired LCU sum.

\end{document}